\pdfoutput=1

\documentclass[11pt,twoside,a4paper,cmspaper,final,collab]{cms-tdr}
\def\svnVersion{231015}\def\cmsCernNo{2013-037}\def\cmsCernDate{\today}\def\cmsMessage{Published in Physics Letters B as \href{http://dx.doi.org/10.1016/j.physletb.2012.12.008}{\doi{10.1016/j.physletb.2012.12.008.}}}

\begin{document}\cmsNoteHeader{EXO-12-001}

\hyphenation{had-ron-i-za-tion}
\hyphenation{cal-or-i-me-ter}
\hyphenation{de-vices}

\RCS$Revision: 159795 $
\RCS$HeadURL: svn+ssh://alverson@svn.cern.ch/reps/tdr2/papers/EXO-12-001/trunk/EXO-12-001.tex $
\RCS$Id: EXO-12-001.tex 159795 2012-12-02 20:57:01Z alverson $
\newlength\cmsFigWidth
\ifthenelse{\boolean{cms@external}}{\setlength\cmsFigWidth{1.0\columnwidth}}{\setlength\cmsFigWidth{0.4\textwidth}}
\ifthenelse{\boolean{cms@external}}{\providecommand{\cmsLeft}{top}}{\providecommand{\cmsLeft}{left}}
\ifthenelse{\boolean{cms@external}}{\providecommand{\cmsRight}{bottom}}{\providecommand{\cmsRight}{right}}
\newlength\cmsFigWidthN
\ifthenelse{\boolean{cms@external}}{\setlength\cmsFigWidthN{1.0\columnwidth}}{\setlength\cmsFigWidthN{0.4\textwidth}}
\cmsNoteHeader{EXO-12-001} % This is over-written in the CMS environment: useful as preprint no. for export versions
\newcommand{\usedLumiWprime}{5.0\fbinv}
\newcommand{\PWprL}{\ensuremath{\cmsSymbolFace{W}_\cmsSymbolFace{L}^{'}}}
\newcommand{\PWprR}{\ensuremath{\cmsSymbolFace{W}_\cmsSymbolFace{R}^{'}}}
\newcommand{\PWprMix}{\ensuremath{\cmsSymbolFace{W}_\cmsSymbolFace{LR}^{'}}}

\title{Search for a \PWpr\ boson decaying to a bottom quark and a top quark in pp collisions at $\sqrt{s}$  = 7\TeV}

\date{\today}

\abstract{Results are presented from a search for a \PWpr\ boson using a dataset
corresponding to 5.0\fbinv of integrated luminosity collected during 2011 by the
CMS experiment at the LHC in pp collisions at $\sqrt{s}=7$\TeV. The \PWpr\ boson
is modeled as a heavy \PW\ boson, but different scenarios for the couplings to
fermions are considered, involving both left-handed and right-handed chiral
projections of the fermions, as well as an arbitrary mixture of the two. The
search is performed in the decay channel ${\PWpr}\to \cPqt\cPqb$, leading to
a final state signature with a single electron or muon, missing transverse
energy, and jets, at least one of which is identified as a b-jet. A \PWpr\ boson
that couples to the right-handed (left-handed) chiral projections of the
fermions with the same coupling constants as the \PW\ is excluded for
masses below 1.85 (1.51)\TeV at the 95\% confidence level. For the first time using
LHC data, constraints on the \PWpr\ gauge couplings for a set of
left-~and right-handed coupling combinations have been placed.
These results represent a significant improvement over previously published limits.}

\hypersetup{%
pdfauthor={CMS Collaboration},%
pdftitle={Search for a W' boson decaying to a bottom quark and a top quark in pp collisions at sqrt(s) = 7 TeV},%
pdfsubject={CMS},%
pdfkeywords={CMS, physics, W'}}

\maketitle %maketitle comes after all the front information has been supplied

\section{Introduction}

New charged massive gauge bosons, usually called $\PWpr$, are predicted by various extensions of the
standard model (SM), for example~\cite{LeftRight,Technicolor,ExtraD,LittleHiggs}.
In contrast to the {\PW} boson, which couples only to left-handed fermions, the couplings of the
{\PWpr} boson may be purely left-handed, purely right-handed, or a mixture of the two, depending on the model.
Direct searches for $\PWpr$ bosons have been conducted in leptonic final states and
have resulted in lower limits for the $\PWpr$ mass of
2.15\TeV~\cite{ATLAS:2011} and 2.5\TeV~\cite{CMS:2012}, obtained at the Large Hadron Collider (LHC)
by the ATLAS and CMS experiments respectively.
CMS has also searched for the process $\PWpr \to \PW$Z using the fully leptonic final states and has excluded $\PWpr$ bosons with masses below 1.14\TeV~\cite{CMS:WZ}.
For $\PWpr$ bosons that couple only to right-handed fermions, the decay to leptons
will be suppressed if the mass of the right-handed neutrino is larger than the mass of
the $\PWpr$ boson. In that scenario, the limits from the leptonic searches do not apply.
Thus it is important to search for $\PWpr$ bosons also in quark final states. Searches for dijet resonances
by CMS~\cite{CMSdijet:2011} have led to the limit $M({\PWpr})>1.5$\TeV.

In this Letter, we present the results of a search for {\PWpr} via the
$\PWpr \to \cPqt\cPqb$ ($\cPqt\cPaqb$ + $\cPaqt\cPqb$) decay channel.  This channel is
especially important because in many models the $\PWpr$ boson is expected to be
coupled more strongly to the third generation of quarks than to the first and second generations.
In addition, it is easier to suppress the multijet background for the decay ${\PWpr}\to \cPqt\cPqb$
than for $\PWpr$ decays to first- and second-generation quarks.
In contrast to the leptonic searches,
the $\cPqt\cPqb$ final state is, up to a quadratic
ambiguity, fully reconstructible, which means that one can search for {\PWpr} resonant mass peaks even in the case of wider {\PWpr}  resonances.

Searches in the ${\PWpr}\to \cPqt\cPqb$ channel at
the Tevatron~\cite{CDF:2009,D0:2010,D0Wprime} and at the LHC by the ATLAS
experiment~\cite{ATLAS:2012} have led to the
limit $M({\PWpr})>1.13$\TeV.
 The SM {\PW} boson and a {\PWpr} boson with non-zero left-handed
coupling strength couple to the same
fermion multiplets and hence would interfere with each other in single-top production~\cite{Simmons:1996ws}.
The interference term
may contribute as much as 5-20\%  of the total rate, depending on the
{\PWpr} mass and its couplings~\cite{Boos:2006xe}.
The most recent \DZERO analysis~\cite{D0Wprime},
in which arbitrary admixtures of left-~and right-handed
couplings are considered, and interference effects are included, sets
a lower limit on the {\PWpr} mass of 0.89 (0.86)~\TeV, assuming purely right-handed
(left-handed) couplings. A limit on the {\PWpr} mass for any combination of left-~and right-handed couplings
is also included.

We present an analysis of events with the final state
signature of an isolated electron, \Pe, or muon, \Pgm,
an undetected neutrino causing an imbalance in
transverse momentum, and jets, at least one of which is identified as a b-jet from the decay chain ${\PWpr} \to \cPqt\cPqb$, $\cPqt \to \cPqb{\PW} \to \cPqb\ell\nu$.  The reconstructed $\cPqt\cPqb$ invariant mass is used to search for {\PWpr} bosons with arbitrary combinations of left-~and right-handed couplings.  A multivariate analysis optimized for {\PWpr} bosons with purely right-handed couplings is also used.
The primary sources of background are \ttbar, \PW+jets, single-top
($\cPqt\PW$, $\cPqs$-~and $\cPqt$-channel production),  $\cPZ/\Pgg^*$+jets, diboson production
($\PW\PW$, $\PW\cPZ$), and QCD multijet events with one jet
misidentified as an isolated lepton.
The contribution of these backgrounds is estimated from simulated event samples after applying correction factors derived from data in control regions well separated from the signal region.

\section{The CMS detector}
\label{sec:CMS}
The Compact Muon Solenoid (CMS) detector comprises a superconducting solenoid providing a uniform magnetic field of 3.8\unit{T}.  The inner tracking system comprises a silicon pixel and strip detector covering $|\eta|<2.4$,
where the pseudorapidity $\eta$ is defined as $\eta=-\ln[\tan (\theta/2) ]$. The polar
angle $\theta$ is measured with respect to the counterclockwise-beam direction (positive $z$-axis) and the
azimuthal angle $\phi$ in the transverse $x$-$y$ plane.
Surrounding the tracking volume, a lead tungstate crystal electromagnetic calorimeter (ECAL)
with fine transverse ($\Delta \eta, \Delta \phi$)
granularity covers the region $|\eta|<3$, and a brass/scintillator hadronic calorimeter
covers $ | \eta | < 5$.
The steel return yoke outside the solenoid is instrumented with gas detectors,
which are used to identify muons in the range $ | \eta | < 2.4$.
The central region is covered by drift tube chambers and the forward region by cathode strip chambers,
each complemented by resistive plate chambers. In addition, the CMS detector has an
extensive forward calorimetry. A two-level trigger system selects the most interesting $\Pp\Pp$ collision
events for physics analysis. A detailed description of the CMS detector can be found elsewhere~\cite{:2008zzk}.

\section{\label{sec:MCmodel}Signal and background modeling}

\subsection{Signal modeling}
The most general model-independent lowest-order
 effective Lagrangian for the interaction of the
{\PWpr} boson with SM fermions~\cite{Sullivan:2002jt} can be written as
\ifthenelse{\boolean{cms@external}}
{
\begin{multline}
\mathcal{L} = \frac{V_{f_if_j}}{2\sqrt{2}} g_w \bar{f}_i\gamma_\mu
\Big[ a^R_{f_if_j} (1+{\gamma}^5) + \\
a^L_{f_if_j}
(1-{\gamma}^5) \Big]{\PWpr}^{\mu} f_j + \text{h.c.} \,,
\end{multline}
}
{
\begin{eqnarray}
\mathcal{L} = \frac{V_{f_if_j}}{2\sqrt{2}} g_w \bar{f}_i\gamma_\mu
\left[ a^R_{f_if_j} (1+{\gamma}^5) + a^L_{f_if_j}
(1-{\gamma}^5) \right]{\PWpr}^{\mu} f_j + \text{h.c.} \,,
\end{eqnarray}
}
where $a^R_{f_if_j}, a^L_{f_if_j}$ are the right- and left-handed couplings of the \PWpr~boson to fermions $f_i$ and $f_j$,
$g_w = e/(\sin \theta_\text{W})$ is the SM weak coupling constant, and  $\theta_\text{W}$ is the Weinberg angle.
If the fermion is a quark, $V_{f_if_j}$ is the Cabibbo-Kobayashi-Maskawa matrix element, and if it is a lepton,
$V_{f_if_j}=\delta_{ij}$ where $\delta_{ij}$ is the Kronecker delta and $i$ and $j$ are the generation numbers. The notation is defined such that for a \PWpr~boson with SM couplings $a^L_{f_if_j}=1$ and $a^R_{f_if_j}=0$.

This effective Lagrangian has been incorporated into
the {\textsc{SingleTop}} Monte Carlo (MC) generator~\cite{Boos:2006af}, which simulates electroweak top-quark production processes based on the
complete set of tree-level Feynman diagrams calculated by the \COMPHEP~\cite{Boos:2004kh} package. This generator is used to simulate
the $\cPqs$-channel {\PWpr} signal including interference with the
standard model {\PW} boson. The complete chain of \PWpr, top quark, and SM {\PW} boson decays
are simulated taking into account finite widths and all spin correlations between resonance
state production and subsequent decay.
The top-quark mass, $M_{\cPqt}$, is chosen to be 172.5\GeV. The CTEQ6.6M parton
distribution functions (PDF) are used and the factorization scale is set to
$M({\PWpr})$. Next-to-leading-order (NLO) corrections are included in the {\textsc{SingleTop}} generator
and normalization and matching between various partonic subprocesses are performed, such that
both NLO rates and shapes of distributions
are reproduced~\cite{Sullivan:2002af,Sullivan:2002jt,Cao:2004ky, Cao:2004ap,Boos:2006xe}.

The \COMPHEP simulation samples of ${\PWpr}$ bosons are generated
at mass values ranging from 0.8 to 2.1\TeV. They are further processed with
 {\PYTHIA}~\cite{Sjostrand:2006za} for  parton fragmentation and hadronization. The
simulation of the CMS detector is performed using \GEANT~\cite{geant}. %GEANTfour
The leading-order (LO) cross section computed by \COMPHEP is then scaled to the
NLO using a $k$-factor of 1.2~\cite{Sullivan:2002jt}.

We generate the following simulated samples of $\cPqs$-channel $\cPqt\cPqb$ production: {\PWprL} bosons that couple only to left-handed fermions ($a^L_{f_i f_j}=1,~a^R_{f_i f_j}=0$), {\PWprR} bosons that couple only to right-handed fermions ($a^L_{f_i f_j}=0,~a^R_{f_i f_j}=1$), and {\PWprMix} bosons that couple equally to both ($a^L_{f_i f_j}=1,~a^R_{f_i f_j}=1$). All ${\PWpr}$  bosons decay to  $\cPqt\cPqb$ final states. We also generate a sample for SM
$\cPqs$-channel $\cPqt\cPqb$ production through an intermediate W boson.
Since ${\PWprL}$ bosons couple to the same fermion multiplets as the SM W boson,
there is interference between SM $\cPqs$-channel $\cPqt\cPqb$ production and
$\cPqt\cPqb$ production through an intermediate {\PWprL}  boson. Therefore, it is not possible to
generate separate samples of  SM $\cPqs$-channel $\cPqt\cPqb$ production and $\cPqt\cPqb$
production through {\PWpr}  bosons that couple to left-handed fermions.
The samples for {\PWprL} and {\PWprMix}  include $\cPqs$-channel $\cPqt\cPqb$
production and the interference. The ${\PWprR}$ bosons
couple to different final-state quantum numbers and therefore there is no interference with
$\cPqs$-channel $\cPqt\cPqb$  production. The  {\PWprR} sample includes  $\cPqt\cPqb$ production only
through {\PWprR} bosons. This sample can then simply be added to the  $\cPqs$-channel $\cPqt\cPqb$
production sample to create a sample that includes all processes for $\cPqs$-channel $\cPqt\cPqb$.

The leptonic decays of ${\PWprR}$ involve a right-handed neutrino
$\nu_R$ of unknown mass.
If $M_{{\nu_R}} > M_{\PWpr}$, ${\PWprR}$ bosons can only decay to
$\Pq^{'}\Paq$ final states.
If $M_{\nu_R} \ll M_{{\PWpr}}$, they can also decay to $\ell\nu$ final states
leading to different branching fractions for
${\PWpr}\rightarrow \cPqt\cPqb$. Table~\ref{tab:xsec_theory} lists the NLO production cross section times branching fraction,
$\sigma (\Pp \Pp \to \PWpr ) B (\PWpr \to \cPqt\cPqb)$.
Here ${\sigma}_L$ is the cross section for  $\cPqs$-channel $\cPqt\cPqb$
production in the presence of a {\PWpr} boson which couples to left-handed fermions,
$(a^L,a^R)=(1,0)$ including $\cPqs$-channel production and interference;
${\sigma}_{LR}$ is the cross section for {\PWpr}  bosons that couple
to left- and to right-handed fermions $(a^L,a^R)=(1,1)$,
including SM $\cPqs$-channel $\cPqt\cPqb$ production and interference;
${\sigma}_R$ is the cross section for $\cPqt\cPqb$ production in the presence of
{\PWpr} bosons that couple only to right-handed fermions $(a^L,a^R)=(0,1)$.
The cross section
for SM  $\cPqs$-channel production,  $(a^L,a^R)=(0,0)$, ${\sigma}_\mathrm{SM}$ is taken to be
$4.63\pm0.07^{+0.19}_{-0.17}$ pb~\cite{s-ch}.

\begin{table*}[!h!tb]
\begin{center}
\small
{\topcaption{
NLO production cross section times branching fraction, $\sigma (\Pp \Pp \to W/\PWpr ) B (W/\PWpr \to \cPqt\cPqb)$, in pb,
for different {\PWpr} boson masses.}
\label{tab:xsec_theory}
        }
\footnotesize
\begin{tabular}{l|c|c|c|c|c|c}
\hline \hline
 $M_{{\PWpr}}$     &    \multicolumn{3}{c|}{$M_{{\nu_R}} \ll M_{\PWpr}$} &  \multicolumn{3}{c}{$M_{\nu_R} >  M_{\PWpr}$} \\ \hline
(TeV)& {$\sigma_R$}   & {${\sigma}_L$}  & {${\sigma}_{LR}$}& {$\sigma_R$}   & {${\sigma}_L$}  & {${\sigma}_{LR}$} \\ \hline
~0.9  & 1.17  & 2.28 & 3.22 & 1.56 & 3.04 & 4.30\\
~1.1 & 0.43  & 1.40 & 1.85 & 0.58 & 1.86 & 2.47\\
~1.3 & 0.17  & 1.20 & 1.39 & 0.23 & 1.60 & 1.85\\
~1.5 & 0.07  & 1.13 & 1.21 & 0.099 & 1.51 & 1.62\\
~1.7 & 0.033 & 1.12 & 1.15 & 0.044 & 1.50 & 1.54\\
~1.9 & 0.015 & 1.11 & 1.13 & 0.020 & 1.49 & 1.51\\
\hline \hline
\end{tabular}
\normalsize
\end{center}
\end{table*}

\begin{figure}[htb]
\begin{center}
\includegraphics[width=\cmsFigWidth]{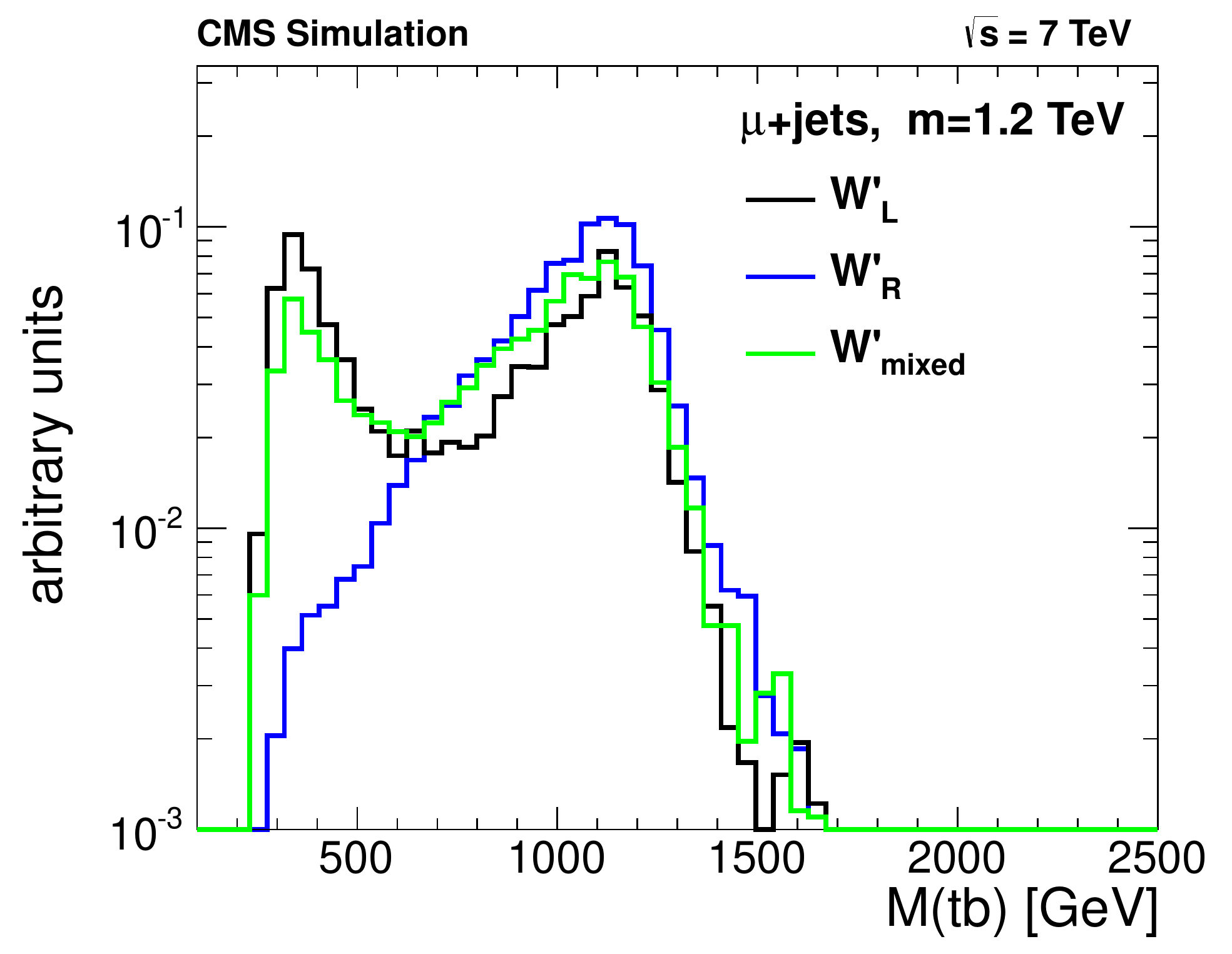}
\end{center}
\caption{Simulated invariant mass distributions for production of ${\PWprR}$,
${\PWprL}$, and $\PWprMix$ with a mass 1.2\TeV.
For the cases of $\PWprL$ and $\PWprMix$, the invariant mass
distributions also include the contribution from $\cPqs$-channel
single top quark production and show a minimum  corresponding to the
destructive interference between the amplitudes for production of left-handed
fermions via the $\PW$ and ${\PWprL}$ bosons. These distributions are after applying the selection criteria  described in Sec.~\ref{sec:selection}.
\label{fig:invMassMC}}
\end{figure}

Figure~\ref{fig:invMassMC} shows the invariant mass distributions for ${\PWprR}$, $\PWprL$,
and $\PWprMix$ bosons. These distributions are obtained after applying
the selection criteria described in Sec.~\ref{sec:selection} and
matching the reconstructed jets, lepton, and an imbalance in transverse momentum of a $\PWpr$ boson with mass 1.2\TeV to the generator level objects. These distributions show a resonant structure around the generated $\PWpr$ mass.
However, the invariant mass distributions for $\PWprL$ and $\PWprMix$ bosons
also include the contribution from s-channel single top quark production and
show a minimum corresponding to the destructive interference between the amplitudes
for production of left-handed fermions via the $\PW$ and $\PWpr$ bosons. The width of a $\PWpr$ boson with a mass of 0.8 (2.1)\TeV is about 25 (80)\GeV, which is smaller than the detector resolution of 10 (13)\%\, and hence does not have an appreciable effect on our search.

\subsection{Background modeling}

Contributions from the background processes are estimated using samples of simulated events.
The $\PW$+jets and Drell--Yan ($\cPZ/\Pgg^* \rightarrow\ell\ell$) backgrounds are
estimated using samples of events generated with the \MADGRAPH 5.1.3~\cite{madgraph} generator.
The \ttbar samples are generated using \MADGRAPH and
normalized to the approximate next-to-NLO (NNLO) cross section~\cite{ttbarNNLOaprox}.
Electroweak diboson ($\PW\PW$,$\PW$Z) backgrounds are generated with {\PYTHIA} and scaled to the NLO
cross section calculated using \MCFM~\cite{MCFM}.
The three single top production channels ($\cPqt\PW$, $\cPqs$-, and $\cPqt$-channel) are estimated using simulated samples generated
with \POWHEG~\cite{powheg}, normalized to the NLO cross section calculation~\cite{s-ch,t-ch,tW-ch}.
For the ${\PWprR}$ search,
the three single-top production channels
are considered as backgrounds. In the analysis  for  ${\PWprL}$ and ${\PWprMix}$ bosons,
because of interference between $\cPqs$-channel single-top production and $\PWpr$,
only $\cPqt\PW$ and $\cPqt$-channel contribute to the backgrounds. Instrumental background due to a jet misidentified as an isolated lepton is estimated using a
sample of QCD multijet background events generated using \PYTHIA. The instrumental background contributions
were also verified using a control sample of multijet events from data.
All  parton-level samples are processed with
\PYTHIA for  parton fragmentation and hadronization and the
response of the detector was simulated using \GEANT.
The samples are further processed through
the trigger emulation and event reconstruction chain of the CMS experiment.

\section{Event selection}
\label{sec:selection}

The $\PWpr \to \cPqt\cPqb$ decay with $\cPqt \to \PW \cPqb$ and $\PW \to \ell$$\nu$ is characterized by the presence of
at least two b jets with high transverse momentum (\pt),
a significant length of the vectorial sum of the negative transverse momenta of all objects in the event (\ETmiss) associated with an escaping neutrino,and a high-\pt isolated lepton.
The isolation requirement is based on the ratio of the total transverse energy observed from
all hadrons and photons in a cone of size $\Delta R = \sqrt{(\Delta\eta)^2+(\Delta\phi)^2} < 0.4$ around the
lepton direction to the transverse momentum of the lepton itself (relative isolation).

Candidate events are recorded if they pass an online trigger requiring
an isolated muon trigger or an electron + jets + \ETmiss trigger and
are required to have at least one reconstructed primary vertex. Leptons, jets, and \ETmiss are
reconstructed using the particle-flow
algorithm~\cite{pf}. At least one lepton is required to be
within the detector acceptance ($|\eta| <2.5$ for electrons excluding the
barrel/endcap transition region, $1.44 < |\eta| < 1.56$, and $|\eta| < 2.1$ for muons).
The selected data samples corresponds to a total integrated luminosity of $5.0 \pm 0.1$\fbinv.

Leptons are required to be separated from jets by $\Delta R(\text{jet},\ell)>0.3$.
Muons are required to have relative isolation less than 0.15 and transverse momentum $\pt > 32\GeV $. The track associated with a muon candidate is
required to have at least ten hits in the silicon tracker,
at least one pixel hit and a good quality global fit with
$\chi^2$ per degree of freedom ${<}10$ including at least one hit in the muon detector.
Electron candidates are selected using shower-shape information, the quality of the track
and the match between the track and electromagnetic cluster, the fraction of total cluster
energy in the hadronic calorimeter, and the amount of activity in the surrounding
regions of the tracker and calorimeters~\cite{eID}.
Electrons are required to have relative isolation less than 0.125, $\pt > 35$\GeV, and
are initially identified by matching a track to a cluster of energy in the ECAL.
Events are removed whenever the electron is determined to
originate from a converted photon.
Events containing a second lepton with relative isolation requirement less than 0.2 and
a minimum $\pt$ requirement for muons (electrons) of 10\GeV (15\GeV) are also rejected.
Additionally, the cosmic-ray background is reduced by requiring
the transverse impact parameter of the lepton with respect to the beam spot to
be less than 0.2\mm.

Jets are clustered using the anti-$k_\mathrm{T}$ algorithm with a size parameter
$\Delta R=0.5$~\cite{antikt} and are required to have $\pt > 30\GeV$ and $|\eta| < 2.4$.
Corrections are applied to account for the dependence of the jet response
as a function of $\pt$ and $\eta$~\cite{JES} and the effects of multiple primary
collisions at high instantaneous luminosity.
At least two jets are required in the event with the
leading jet $\pt > 100\GeV$ and second leading jet $\pt > 40\GeV$. Given that there would be two b quarks
in the final state, at least one of the two leading jets is required to be tagged as a b jet.
Events with more than one b-tagged jet are allowed.
The combined
secondary vertex tagger~\cite{CMS-PAS-BTV-11-004} with the medium operating point is used for this analysis.
The chosen operating point is found to provide best sensitivity based on signal acceptance and expected limits~\cite{CMS-PAS-BTV-11-003}.

The QCD multijet background is reduced by requiring $\ETmiss > 20\GeV$ for the muon + jets channel. Since the multijet
background from events in which a jet is misidentified as a lepton is larger for the electron + jets channel, and because
of the presence of a \ETmiss requirement in the electron trigger, a tighter $\ETmiss > 35$\GeV requirement is imposed for this channel.

To estimate the \PWpr\ signal and background yields, data-to-MC scale factors (g)
measured using Drell--Yan data are applied in order to account for the differences
in the lepton trigger and in the identification and isolation efficiencies.
Scale factors related to the b-tagging efficiency
and the light-quark tag rate (misidentification rate), with a jet $\pt$ and $\eta$ dependency, are applied on a
jet-by-jet basis to all b-, c-, and light quark jets
in the various MC samples~\cite{CMS-PAS-BTV-11-003}.

Additional scale factors are applied to \PW+jets events in which a b quark,
a charm quark, or a light quark is produced in association with the {\PW} boson.
The overall \PW+jets yield is  normalized to the NNLO cross section~\cite{EWKsigma} before requiring a b-tagged jet.
The fraction of heavy flavor events (\PW$\cPqb\cPaqb$, \PW$\cPqc\cPaqc$)  is scaled by
an additional empirical correction  derived using lepton+jets samples with various
jet multiplicities~\cite{CMS-PAS-TOP-11-003}.
Since this correction was obtained for events with a different topology than those
selected in this analysis, an additional correction factor is derived using two data samples:
 events containing zero b-quark jets (0-b-tagged sample)
and the inclusive sample after all the selection criteria, excluding any b-tagging
requirement (preselection sample).
Both samples are background dominated with negligible signal contribution.
 By comparing the \PW+jets background prediction with observed data
in these two samples, through an iterative process, we extract  \PW+light-flavor jets
(g$_{Wlf}$) and \PW+heavy-flavor jets (g$_{Whf}$) scale factors.
The value of the \PW+heavy-flavor jets scale factor determined via this method
is within the uncertainties of the g$_{Whf}$ corrections derived in Ref.~\cite{CMS-PAS-TOP-11-003}.
Both g$_{Wlf}$  and  g$_{Whf}$ scale factors are applied to obtain the expected number of $\PW$+jets events.

The observed number of  events and the expected background yields after applying the above selection criteria and scale factors
are listed in Table~\ref{tab:yields}. These numbers are in agreement between
the observed data and the expected background yields.
The signal efficiency ranges from 87\% to 67\% for $\PWprR$ masses from
0.8 to 1.9\TeV respectively.

\begin{table*}[!h!tb]
\begin{center}
\small
      \topcaption{
        Number of events observed, and number of signal and background  events predicted.
For the background samples, the expectation is computed
corresponding to an integrated luminosity of $\usedLumiWprime$.
The total background yields include the normalization uncertainty on the predicted backgrounds.
``Additional selection'' corresponds to requirements
of the \PWpr invariant mass analysis (described in Sec.~\ref{sec:invmass}) and are:
$\pt(\text{top})>75\GeV$, $\pt(\text{jet1,jet2})>100\GeV$, $130<M(\text{top})<210\GeV$.
\label{tab:yields}
        }
\footnotesize
\begin{tabular}{l|c|c|c|c|c|c}
\hline \hline
    & \multicolumn{6}{c}{Number of events} \\ \hline
Process &  \multicolumn{3}{c|}{ $\Pe$+jets }  & \multicolumn{3}{c}{ $\Pgm$+jets } \\ \hline
  &  \multicolumn{2}{c|}{b-tagged jets} & Additional & \multicolumn{2}{c|}{b-tagged jets}  &  Additional \\
\textbf{Signal}                                      &  =1   & $\geq 1$   & selection  & = 1    & $\geq 1$  & selection  \\ \hline
${\PWprR}$ (0.8\TeV)                  & 405  & 631       &  463       & 539    &  838    &  605    \\
${\PWprR}$ (1.2\TeV)                  & 63  & 90         &  68        & 76    &  109    &  81    \\
${\PWprR}$ (1.6\TeV)                  & 11  & 14         &  11        & 11    &  15    &  11    \\
${\PWprR}$ (1.9\TeV)                  & 3  & 4           &  3         & 3    &  4    &  3    \\ \hline															
\multicolumn{7}{l}\textbf{Background}     \\ \hline	  			       							\ttbar                            & 8496 & 10659      & 4795      & 13392  & 16957  &  6692 \\
$\cPqt$-channel                               & 587  & 686        & 300       & 1047   & 1223   &  442  \\
$\cPqs$-channel                               & 46  &           73                &  32        & 81    &  134   &  51    \\
$\cPqt{\PW}$-channel                           & 549            & 628              & 270       & 886    & 1007   &  395  \\
${\PW}(\rightarrow)\ell\nu$+jets           & 4588          & 4760          & 1404         & 8673   & 9023   &  2350 \\
Z$\gamma^*(\rightarrow\ell\ell$)+jets & 164                & 173             & 68          & 388    & 414    &  135  \\
Diboson                                   & 51            & 52              & 17           & 77     & 79     &  27   \\
Multijet QCD                    & 104            & 225              &  0           & 121    & 121    &  0    \\ \hline \hline
{\bf Total background}                    & 14585$\pm$3199 & 17256$\pm$3780 & 6886$\pm$1371  & 24665$\pm$4917  & 28958$\pm$5765  &  10092$\pm$1807 \\ \hline\hline
{\bf Data}                             & 14337            & 16758           &   6638        & 23979 & 28392  & 9821   \\ \hline \hline
\end{tabular}
\normalsize
\end{center}
\end{table*}

\section{Data analysis}

In this section, we describe two analyses to search for $\PWpr$ bosons.
The reconstructed $\cPqt\cPqb$ invariant mass analysis is used to search
for {\PWpr} bosons with arbitrary combinations of left-~and right-handed couplings
while a multivariate analysis is optimized for the search of {\PWpr} bosons with
purely right-handed couplings.

\subsection{The \texorpdfstring{$\cPqt\cPqb$}{tb} invariant mass analysis}
\label{sec:invmass}

The distinguishing feature of a \PWpr signal is a
resonant structure in the $\cPqt\cPqb$ invariant mass.  However, we cannot
directly measure the $\cPqt\cPqb$ invariant mass. Instead we reconstruct the
invariant mass from the combination of
the charged lepton, the neutrino, and the jet that gives the
best top-quark mass reconstruction, and the highest $\pt$ jet that is not associated with the top-quark.
The \ETmiss is used to obtain the $xy$-components of the neutrino momentum. The
$z$-component is calculated by constraining the  \ETmiss and lepton momentum to the $\PW$-boson mass (80.4\GeV).
This constraint leads to a quadratic equation in  $|p_z^{\nu}|$. When the $\PW$ reconstruction
yields two real solutions, both solutions are used
to reconstruct the top candidates. When the solution is complex,
the \ETmiss is minimally modified to give one real solution.
In order to reconstruct the top quark momentum vector,
the neutrino solutions are used to compute the possible $\PW$ momentum vectors.
The top-quark candidates are then reconstructed using the
possible $\PW$ solutions and all of the selected jets in the event. The
candidate with mass closest to 172.5\GeV is chosen as the best
representation of the top quark ($M(\PW,\text{best jet})$). The {\PWpr} invariant mass ($M(\text{best jet, jet2}, \PW)$)
is obtained by combining the ``best''  top-quark candidate
with the highest $\pt$ jet (jet2) remaining after the top-quark reconstruction.

\begin{figure}[htb]
\begin{center}
\includegraphics[width=\cmsFigWidth]{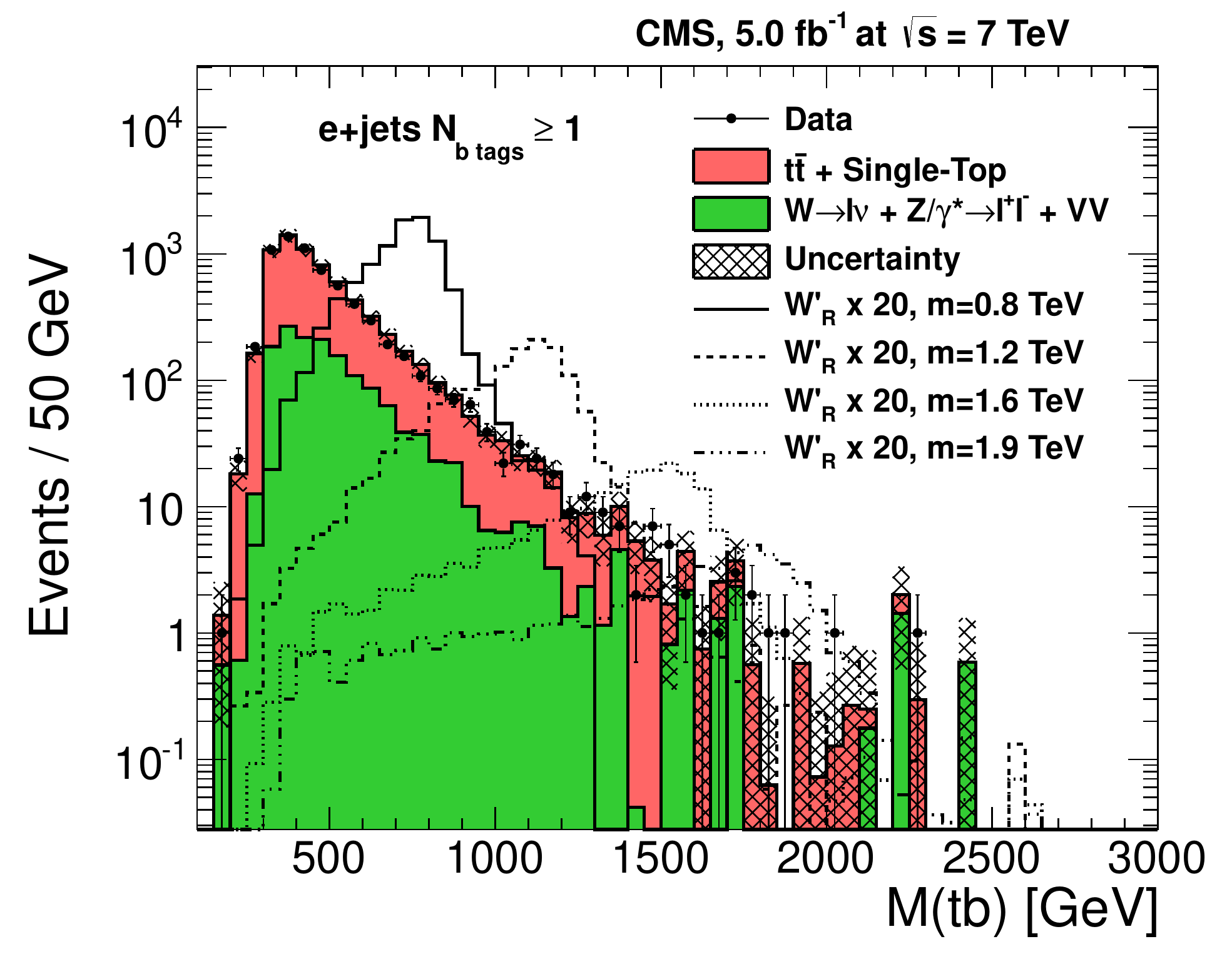}
\includegraphics[width=\cmsFigWidth]{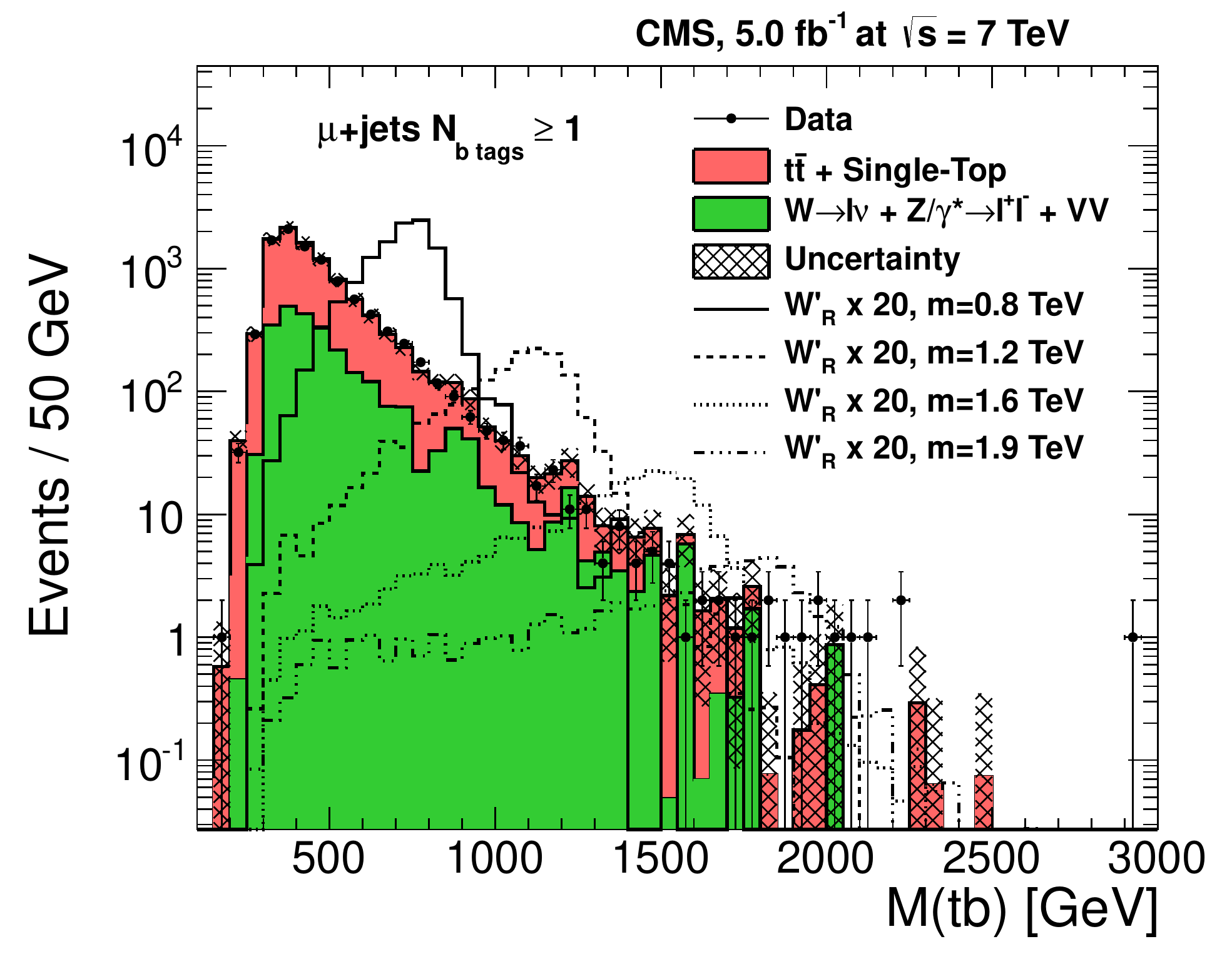}
\end{center}
\caption{
Reconstructed {\PWpr} invariant mass distributions after the full selection. Events with electrons
(muons) are shown in the \cmsLeft\ panel (\cmsRight\ panel) for data, background, and four
different $\PWprR$ signal mass points (0.8, 1.2, 1.6, and 1.9\TeV).
The hatched bands represent the total normalization uncertainty in the predicted backgrounds.
For the purpose of illustration, the expected yields for \PWpr signal samples are scaled by a factor of 20.
\label{Fig:data-bkg}}
\end{figure}

Figure~\ref{Fig:data-bkg} shows the reconstructed $\cPqt\cPqb$ invariant mass distribution
for the data and simulated {\PWpr} signal samples generated at four different mass
values (0.8, 1.2, 1.6, and 1.9\TeV). Also included in the plots are the main background
contributions. The data and background distributions are shown for sub-samples with
one or more b tags, separately for the electron and muon channels.
Three additional criteria are used in defining
the $\geq 1$ b-tagged jet sample to improve
the signal-to-background discrimination: the $\pt$ of the best top candidate must be greater than
$75$\GeV, the $\pt$ of the system comprising of the two leading jets
$\pt(\text{jet1,jet2})$ must be greater than $100$\GeV, and the best top candidate must have
a mass $M(W,\text{best jet})$ greater than 130\GeV and less than 210\GeV.

Since the $\PW$+jets process is one of the major backgrounds to the {\PWpr} signal
(see Table~\ref{tab:yields}), a study is performed to verify that the $\PW$+jets shape
is modeled realistically in the simulation. Events
with zero b-tagged jets in data that satisfy all other selection criteria are
expected to originate predominantly from the $\PW$+jets background. These events
are used to verify the shape of the $\PW$+jets background invariant mass distribution in data.
The shape is obtained by subtracting the backgrounds other than $\PW$+jets from the data.
The invariant mass distribution with zero b-tagged jets derived from data using this method is compared with
that from the W+jets MC sample. They were found to be in agreement, validating the simulation.
Any small residual difference is taken into account as a systematic uncertainty. The difference between
the distributions is included as a systematic uncertainty on the shape of the
$\PW$+jets background. Using MC samples, it was also checked that the shape of
$\PW$+jets background does not depend on the number of b-tagged jets by
comparing the  $\cPqt\cPqb$ invariant mass distribution with and without b-tagged jets with
the distribution produced by requiring one or more b-tagged jets.

\subsection{The boosted decision tree analysis}
The boosted decision tree (BDT) multivariate analysis technique~\cite{BDT, BDT1, BDT2} is also
used to distinguish  between the ${\PWpr}$ signal and the background. For the BDT analysis we apply all the selection criteria described in Sec.~\ref{sec:selection}, except the additional selection given in Table~\ref{tab:yields}.
This method, based on judicious selection of discriminating variables, provides a considerable
increase in sensitivity for the {\PWpr} search compared to the $\PWpr$ invariant mass analysis,  described in Sec.~\ref{sec:invmass}.

The discriminating variables used for the BDT analysis fall into the following categories: object kinematics such as
individual transverse momentum ($\pt$) or pseudorapidity ($\eta$) variables;  event kinematics,
e.g. total transverse energy or invariant mass variables; angular correlations,
either $\Delta R$, angles $\Delta \phi$ between jets and leptons, or top-quark spin correlation
variables; and top-quark reconstruction variables identifying which jets to use for the top quark
reconstruction. The final set of variables chosen for this analysis is shown in Table~\ref{tab:variables}. The ``jet$_{1,2,3,4}$'' corresponds to first, second, third and fourth highest $\pt$ jet; ``btag$_{1,2}$''
corresponds to first, second highest $\pt$ b-tagged jet; ``notbest$_{1,2}$'' corresponds to highest
and second highest \pt jet not used in the reconstruction of best top candidate. Class ``alljets''
includes all the jets in the event in the global variable.
The sum of the transverse energies is $H_T$. The invariant mass of the objects is $M$.
The transverse mass of the objects is $M_T$. The sum of z-components of the momenta of all jets is $p_z$.
The angle between x and y, is $\cos$(x,y)$_{r}$ where the subscript indicates the reference frame.

\begin{table}
\begin{center}
\small
\caption{Variables used for  the multivariate analysis in four different categories. For the angular variables, the subscript indicates the reference frame.
\label{tab:variables}  }
\begin{tabular}{l|l}
\hline
\hline
 {\bf{Object kinematics}}                       & {\bf{Event kinematics}}                 \\
~~$\eta$(jet1)				        &~~Aplanarity(alljets)                    \\
~~$\pt$(jet1)                                &~~Sphericity(alljets)                    \\
~~$\eta$(jet2)                                  &~~Centrality(alljets) 		          \\
~~$\pt$(jet2)                                   &~~$M$(btag1,btag2,$\PW$)	                  \\
~~$\eta$(jet3)                                 	&~~$M$(jet1,jet2,$\PW$)                     \\
~~$\pt$(jet3)                                   &~~$M$(alljets)       	                  \\
~~$\eta$(jet4)                                  &~~$M$(alljets,$\PW$)                        \\
~~$\eta$(lepton)                                &~~$M$($\PW$)    	                          \\
~~$\pt$(lightjet)                               &~~$M$(alljets,lepton,{\ETmiss}) 	   \\
~~$\pt$(lepton)                                 &~~$M$(jet1,jet2)	    	           \\
~~$\eta$(notbest1)                             	&~~$M_T$($\PW$)   	                    \\
~~$\pt$(notbest1)                               &~~$\pt$(jet1,jet2)		    	  \\
~~$\pt$(notbest2)                               &~~$\pt$(jet1,jet2,$\PW$)	                   \\
~~{\ETmiss}                                     &~~$p_z$/$H_T$(alljets)                    \\
{\bf{Top quark reconstruction}}                 &{\bf{Angular correlations}}                 \\
~~$M(W,{\rm btag1})$ (``btag1'' top mass)        & ~~$\Delta \phi$(lepton,jet1)	               \\
~~$M(W,{\rm best1})$ (``best'' top mass)        & ~~$\Delta \phi$(lepton,jet2)	             \\
~~$M(W,{\rm btag2})$ (``btag2'' top mass)     & ~~$\Delta \phi$(jet1,jet2)                \\
~~$\pt (W,{\rm btag1})$ (``btag1'' top \pt)     & ~~$\cos$(best,lepton)$_{\rm besttop}$     \\
~~$\pt (W,{\rm btag2})$ (``btag2'' top \pt)     & ~~$\cos$(light,lepton)$_{\rm besttop}$    \\
                                              	& ~~$\Delta R$(jet1,jet2)                 \\
\hline\hline
\end{tabular}
\normalsize
\end{center}
\end{table}

The input variables selected for the BDT are checked for accurate modeling.
We consider an initial set of
about 50 variables as inputs to the BDT. The selection of the final list
of input variables uses important components from the BDT training procedure,
namely the  ranking of variables in the order of their importance and
correlations among these variables.  In order to  maximize the
information and keep the training optimal, the variables with
smallest correlations are selected. The final list of variables is determined
through an iterative process of
training and selection (based on ranking and correlations), and
the degree of agreement between the data and MC in
two background-dominated regions ($\PW$+jets and \ttbar).
While the relative importance of the various variables used by
the BDT depends on the $\PWpr$ mass, for a 2\TeV $\PWprR$,
the four most important variables are
$\cos$(best,lepton)$_\text{besttop}$, $M$(alljets), $\Delta \phi$(lepton,jet1), and
$\pt$(jet1).
The $\PW$+jets dominated sample is defined by requiring exactly two jets, at least one b-tagged jet, and the scalar sum of the transverse energies of all kinematic objects in the event to be less than  300\GeV. The  \ttbar dominated sample is defined by requiring more than  four jets, and at least one b-tagged jet.

\begin{figure}[htb]
  \begin{center}

     \includegraphics[width=\cmsFigWidth]{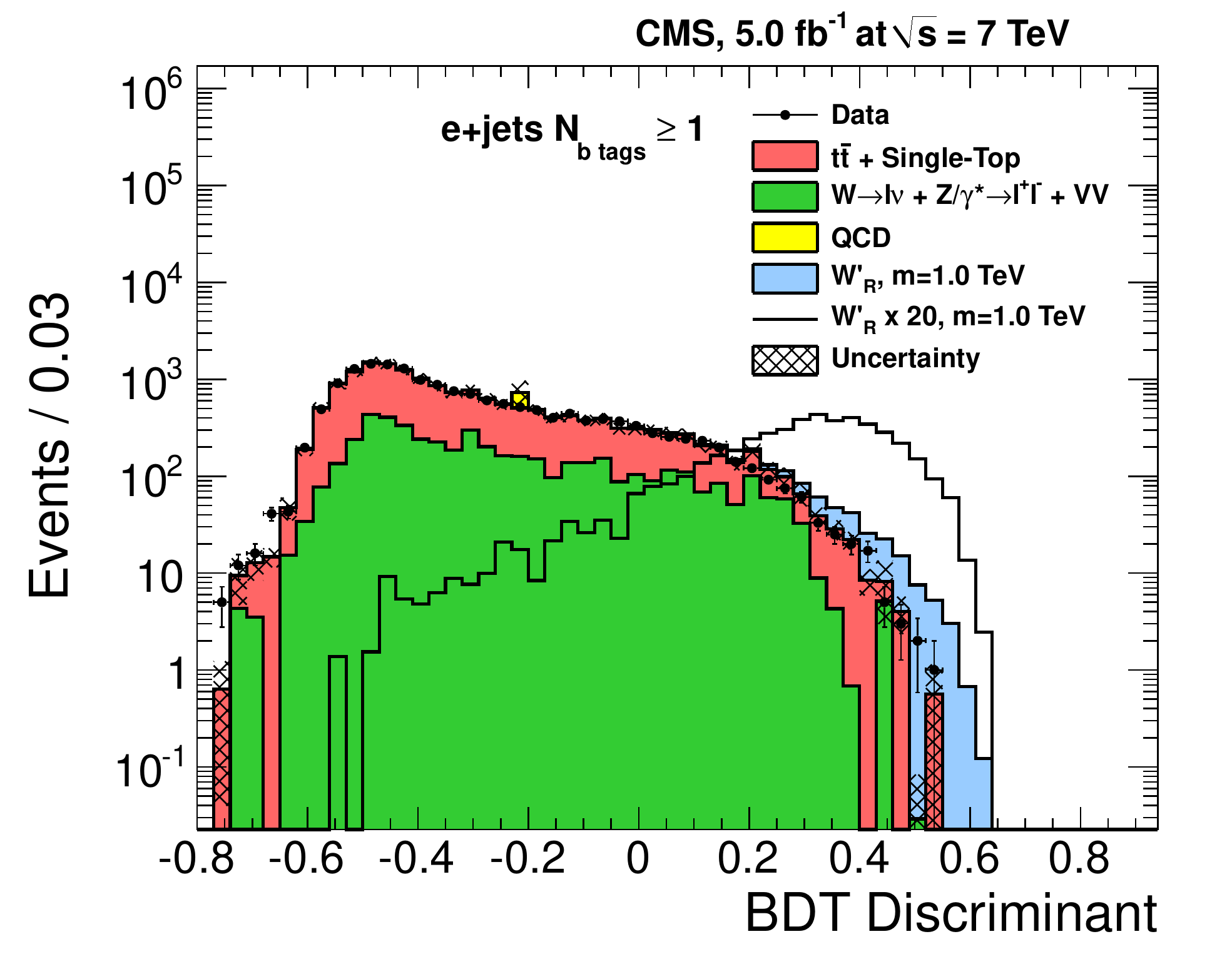}
     \includegraphics[width=\cmsFigWidth]{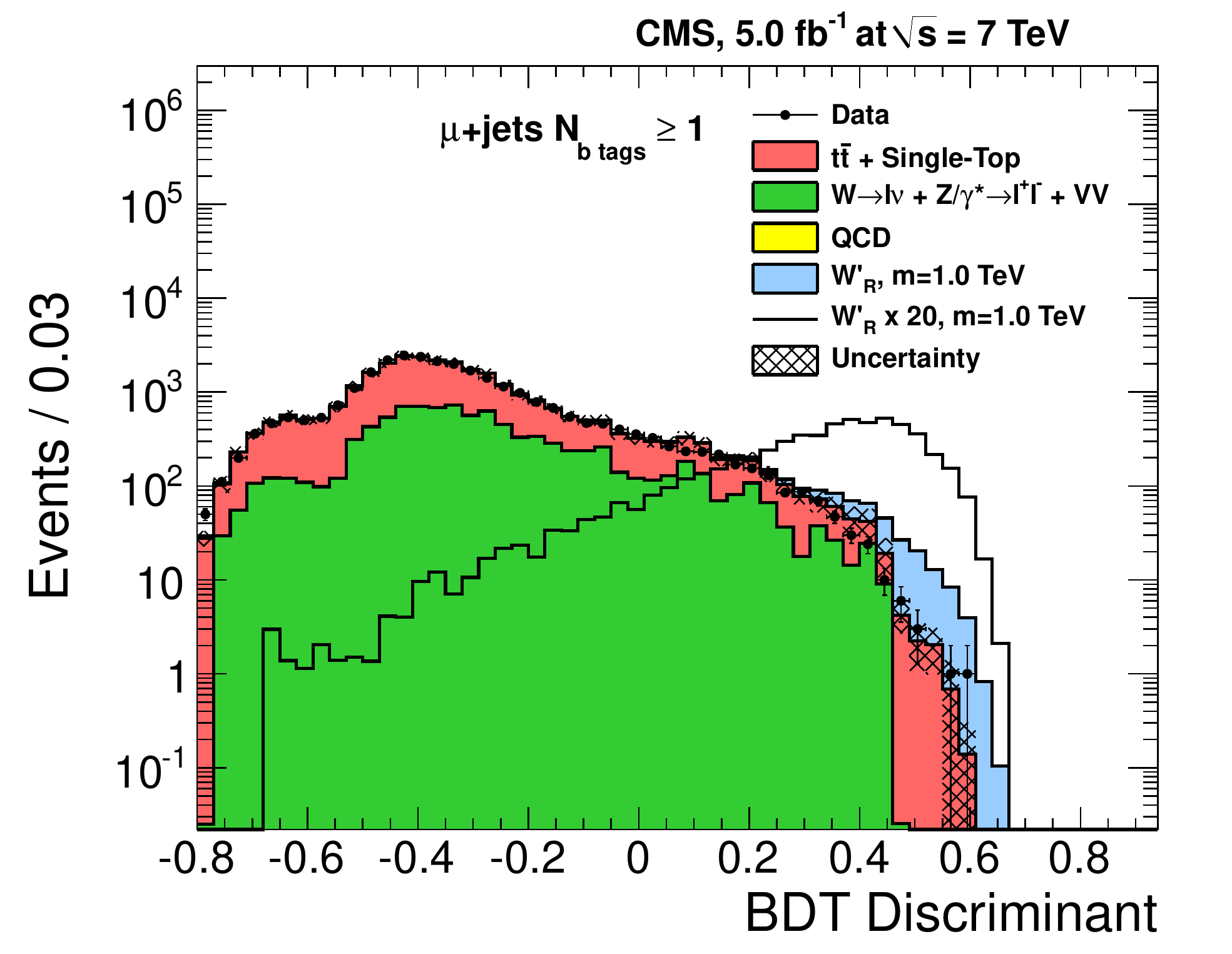}

     \caption{
Distribution of the BDT output discriminant. Plots for the $\Pe$+jets (\cmsLeft) and
the $\Pgm$+jets (\cmsRight) samples are shown for data, expected backgrounds, and a
$\PWpr_R$ signal with  mass of 1\TeV.
The hatched bands represent the total normalization uncertainty on the predicted backgrounds.
    \label{fig:BDT-GE1BTag}
  }
 \end{center}
\end{figure}

The BDTs are trained at each  {\PWpr} mass.
We use the Adaptive Boost Algorithm (AdaBoost) with value 0.2 and 400 trees for training. We use
the Gini index~\cite{Gini} as the criterion for node splitting.
The training to distinguish between signal and the total expected background
is performed separately for the electron and muon event samples, after
requiring the presence of one or more b-tagged jets.
In order to avoid training bias, the background and signal samples are split into
two statistically independent samples. The first sample is used for training of the BDT and
the second sample is used to obtain the final results for the $\PWpr$ signal expectations.
Cross checks are performed by comparing the data and MC for various
BDT input variables and the output discriminants in two control regions, one dominated by
$\PW$+jets background events and the other by $\ttbar$ background events. Figure~\ref{fig:BDT-GE1BTag} shows
data and background comparison for a ${\PWprR}$ with mass of 1\TeV, for both $\Pe$+jets and $\Pgm$+jets events.

\section{\label{sec:sys}Systematic uncertainties}

The sources of systematic uncertainties fall into two categories: (i) uncertainties in the normalization,
and (ii) uncertainties affecting both shape and normalization of the distributions. The first category includes uncertainties
on the integrated luminosity (2.2\%)~\cite{CMS-PAS-SMP-12-008},
theoretical cross-sections and branching fractions (15\%), object identification efficiencies (3\%),
and trigger modeling (3\%).
The uncertainty in the \PWpr~cross section is
about 8.5\% and includes contributions from the NLO scale (3.3\%), PDFs (7.6\%),
$\alpha_s$ (1.3\%), and the top-quark mass ($<1$\%).
Also included in this group are uncertainties related to
obtaining the heavy-flavor ratio from data~\cite{CMS-PAS-TOP-11-003}.
In the limit estimation, these are defined through log-normal priors based on their
mean values and their uncertainties.
The shape-changing category includes the uncertainty from the jet energy scale,
the b-tagging efficiency and misidentification rate scale factors.
For the \PW+jets samples, uncertainties on the light- and heavy-flavor scale
factors  are also included. This uncertainty has the largest impact in the limit estimation.
The variation of the factorization scale $Q^2$ used in the
strong coupling constant $\alpha_s(Q^2)$, and the jet-parton matching scale~\cite{MLM}
uncertainties are evaluated for the \ttbar background sample.
In the case of \PW+jets, there is an additional systematic uncertainty due to the
shape difference between data and simulation as
observed in the 0-b-tagged sample.
These shape uncertainties are evaluated by raising and lowering the corresponding correction by one
standard deviation and repeating the complete analysis.
Then, a bin-wise interpolation using a cubic spline
between histogram templates at the different variations is performed.
A nuisance parameter is associated to the interpolation and included in the limit
estimation. Systematic uncertainties from a mismodeling of the number
of simultaneous primary interactions is found to be negligible in this analysis.

\section{Results}

The observed $\PWpr$ mass distribution (Fig.~\ref{Fig:data-bkg}) and the BDT discriminant distributions
(Fig.~\ref{fig:BDT-GE1BTag}) in the data agree with the prediction
for the total expected background within uncertainties.
We proceed to set upper limits on the {\PWpr} boson production cross section for
different {\PWpr} masses.

\subsection{Cross section limits}

The limits are computed using a variant of the CL$_\mathrm{s}$ statistic~\cite{CLs1,CLs2}.
A binned likelihood is used to calculate upper limits on the signal
production cross section times branching fraction:
$\sigma (\Pp \Pp \to \PWpr ) B (\PWpr \to \cPqt\cPqb \to \ell\nu\cPqb\cPqb)$.
The procedure accounts for the effects on normalization and shape from systematic
uncertainties, see Sec. ~\ref{sec:sys}, as well as
for the limited number of events in the background templates.
Expected cross section limits for each {$\PWprR$} boson mass are also computed
as a measure of the sensitivity of the analysis.
To obtain the best sensitivity, we combine the muon and electron samples.

\begin{figure*}[htb]
  \begin{center}
     \includegraphics[width=\cmsFigWidthN]{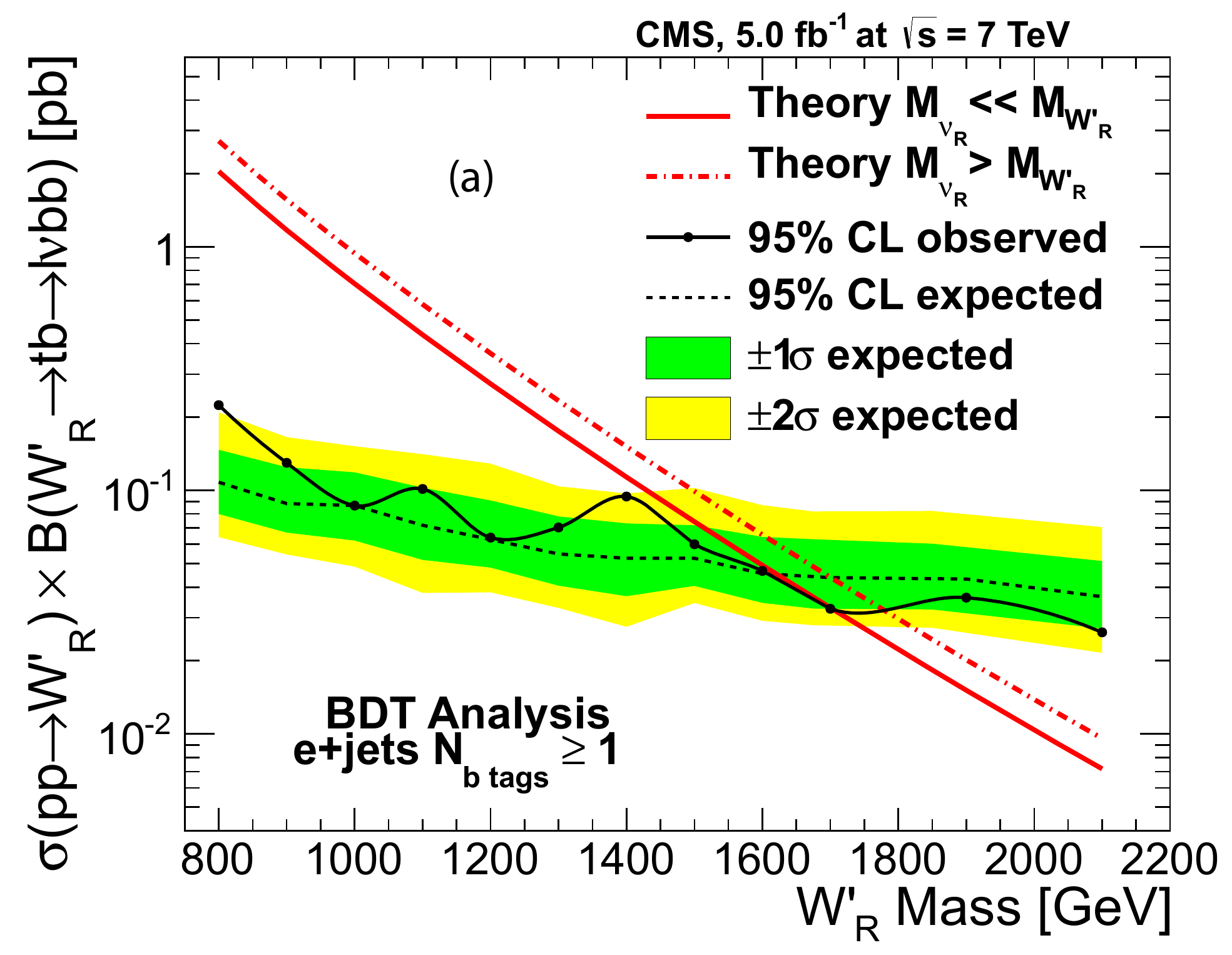}
      \includegraphics[width=\cmsFigWidthN]{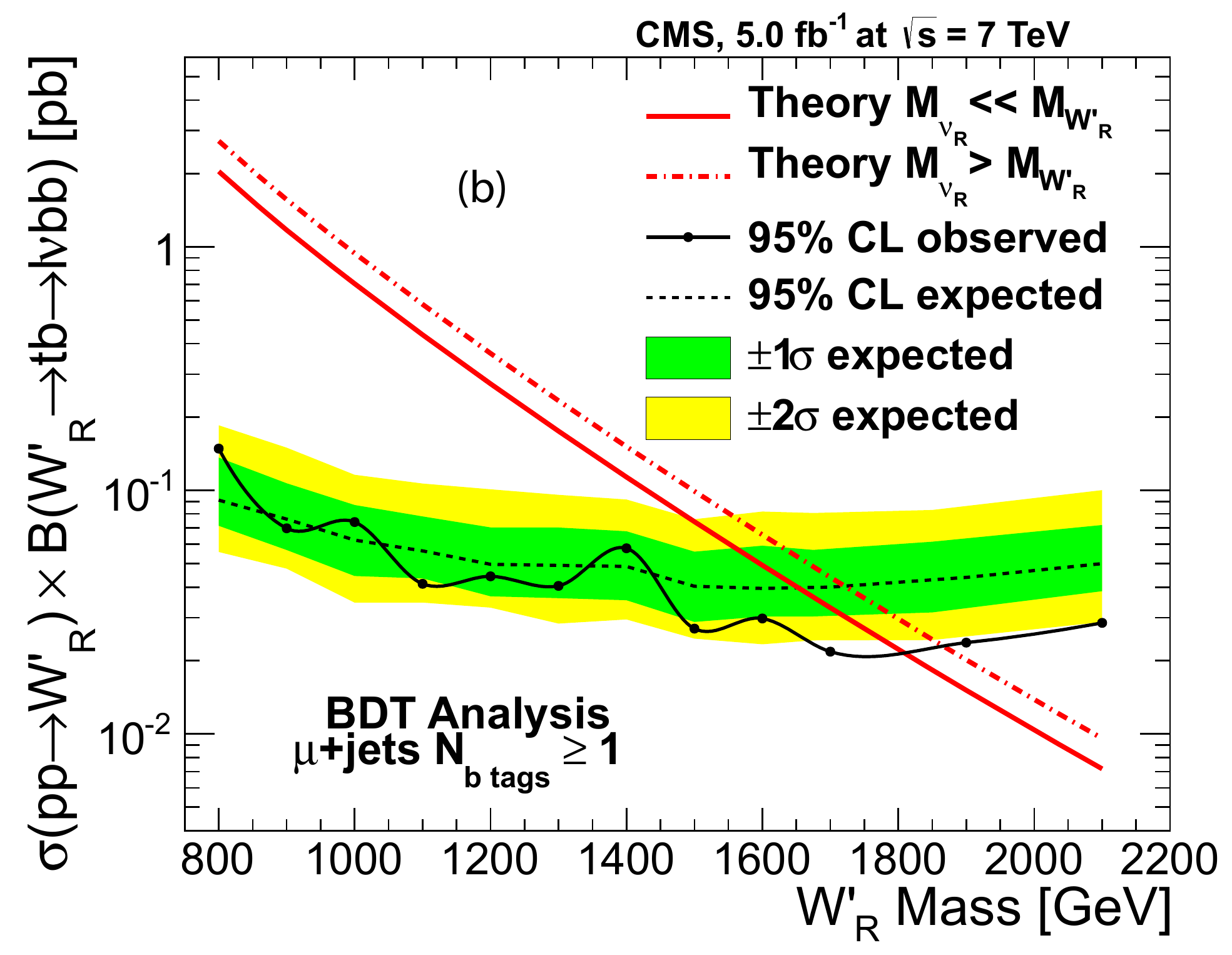}
     \includegraphics[width=\cmsFigWidthN]{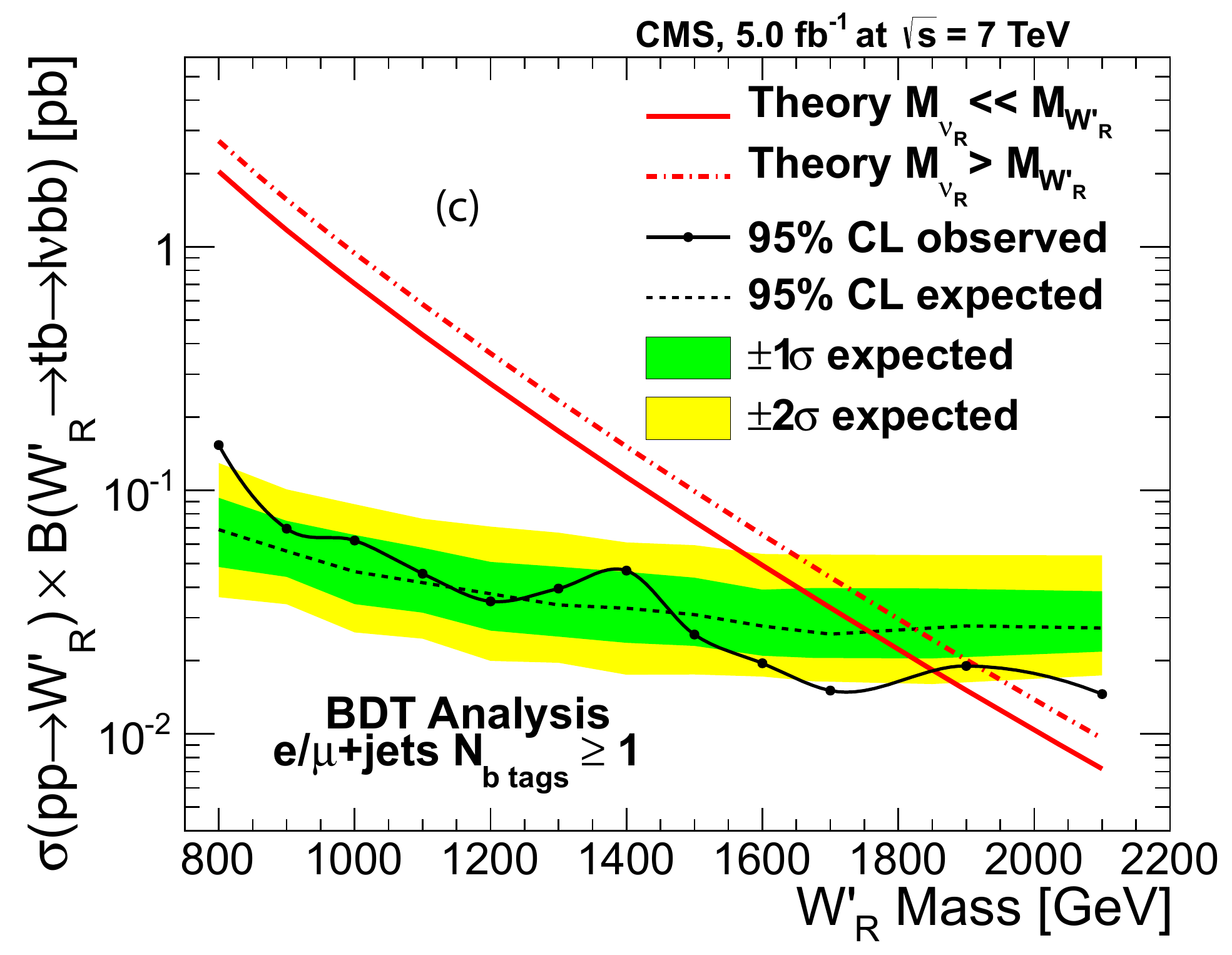}
    \includegraphics[width=\cmsFigWidthN]{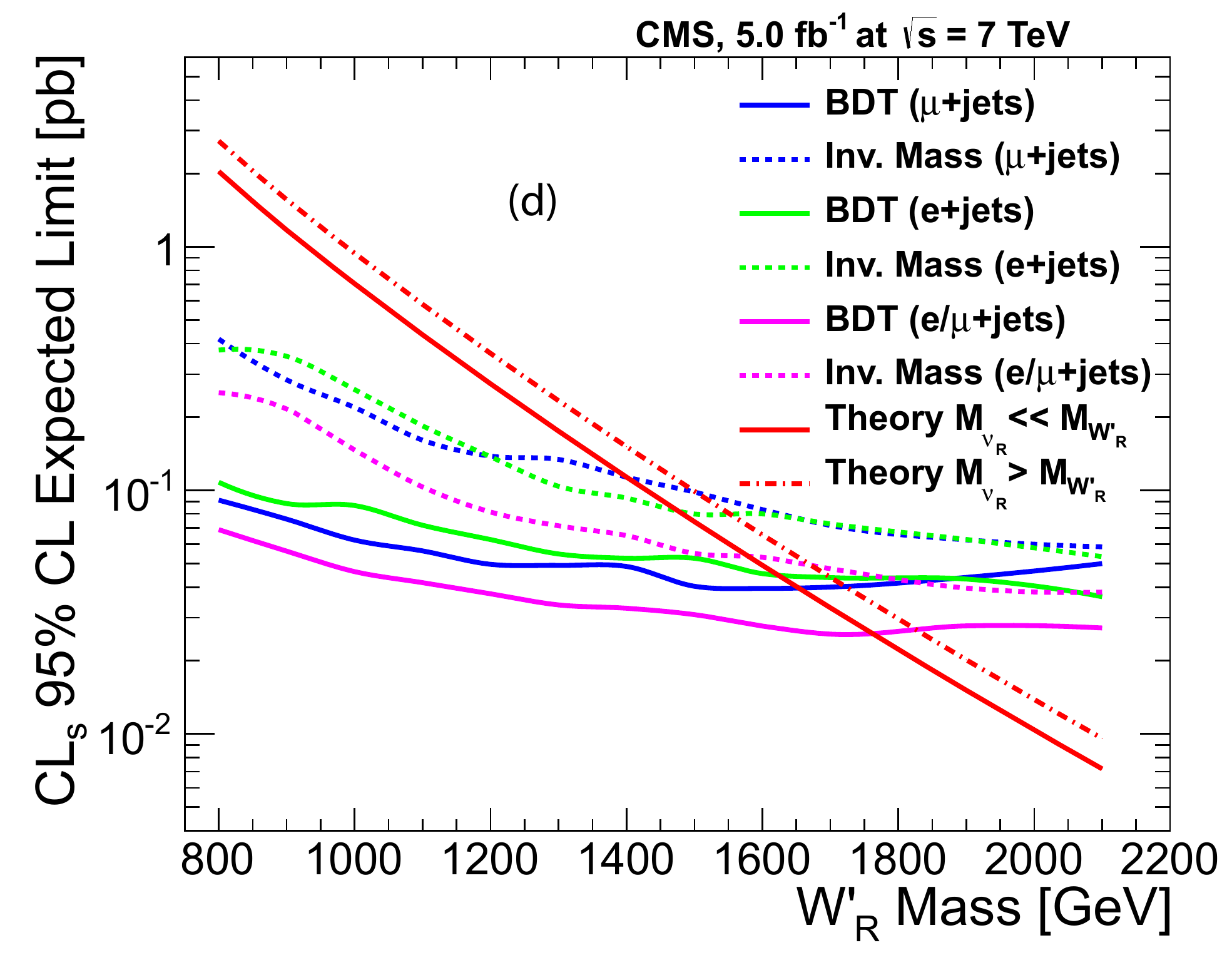}
    \caption{The expected and measured 95\% CL upper limits on the production cross section
$\sigma (\Pp \Pp \to \PWpr ) B (\PWpr \to \cPqt\cPqb \to \ell\nu\cPqb\cPqb)$
of right handed  {\PWpr} bosons obtained using the BDT discriminant for  $\geq$ 1 b-tagged
electron+jets events (a), muon+jets events (b), and combined (c).
Also shown  (d) is a comparison of
the expected 95\% CL upper cross section limits obtained using invariant mass
distribution and BDT output for right handed {\PWpr} bosons  for $\geq$ 1 b-tagged muon+jet events,
electron+jet events, and combined.
The $\pm 1 \sigma$ and $\pm 2 \sigma$ excursions from expected limits are also shown.
The solid and dot-dashed red lines represent the theoretical
cross section predictions for the two scenarios $M_{{\nu_R}} > M_{\PWpr}$, where ${\PWpr}$ can decay only to quarks
and  $M_{\nu_R} \ll M_{{\PWpr}}$, where all decays of ${\PWpr}$ are allowed.~\cite{Sullivan:2002jt,Boos:2006af,Boos:2004kh}.
  \label{fig:limit-right}}
 \end{center}
\end{figure*}

The BDT discriminant distributions, trained for every
mass point, are also used to set upper limits on the production
cross section of the ${\PWprR}$. The expected and
measured 95\% CL upper limits on the production cross section times decay branching fraction for the $\PWprR$ bosons
are shown in Fig.~\ref{fig:limit-right}.
The sensitivity achieved using the BDT output discriminant
is greater than that obtained using the shape of the distribution of the {\PWpr} boson invariant mass.

In all the plots shown in Fig.~\ref{fig:limit-right}, the black solid line denotes the
observed limit and the red solid line and dot-dashed lines represent the theoretical
cross section predictions for the two scenarios $M_{{\nu_R}} > M_{\PWpr}$, where ${\PWpr}$ can decay only to quarks
and  $M_{\nu_R} \ll M_{{\PWpr}}$, where all decays of ${\PWpr}$ are allowed.

We define the lower limit on the ${\PWpr}$ mass by the point where the
measured cross section limit crosses the theoretical cross section curves~\cite{Sullivan:2002jt,Boos:2006xe}.
The observed lower limit on the mass of the {\PWpr} boson with purely right-handed coupling to fermions
is listed in Table~\ref{tab:masslim}.

In the electron
channel, we observe 2 events with a mass above 2\TeV with an expected background of $3.0\pm 1.5$ events.
In the muon channel, we observe 6 events with
an expected background of $1.4\pm 0.9$ events.
This gives a total of 8 events with an expected background of $4.4\pm1.7$ events
with a mass above 2\TeV.
The significance of the excursion in the muon channel is 2.2 standard deviations.
The dominant contributions to the expected background above 2\TeV come from $\PW$+jets and top-quark production.

\subsection{Limits on coupling strengths\label{Shat-generalCoupling}}

From the effective Lagrangian given in Eq.~(1), it can be shown that
the cross section for single-top quark production
in the presence of a $\PWpr$ boson can be expressed, for arbitrary
combinations of left-handed ($a^L$) or right-handed ($a^R$) coupling
strengths, in terms of four cross sections,
${\sigma}_L$, ${\sigma}_R$, ${\sigma}_{LR}$, and ${\sigma}_\mathrm{SM}$ of
the four simulated samples, listed in Table~\ref{tab:xsec_theory}, as

\begin{align}\label{eq:xsec}
{\sigma} &= {\sigma}_\mathrm{SM} + a^L_{ud}a^L_{tb}
\left({\sigma}_L - {\sigma}_R - {\sigma}_\mathrm{SM} \right)  \nonumber \\
         &+ \left(\left(a^L_{ud} a^L_{tb}\right)^2
          +        \left(a^R_{ud} a^R_{tb}\right)^2\right) {\sigma}_R \nonumber \\
         &+ \frac{1}{2}\left(\left(a^L_{ud} a^R_{tb}\right)^2
          +                   \left(a^R_{ud} a^L_{tb}\right)^2\right)
\left( {\sigma}_{LR} - {\sigma}_L - {\sigma}_R  \right).
\end{align}

We assume that the couplings to first-generation quarks, $a_{ud}$, which are important for the production of the
$\PWpr$ boson, and the couplings to third-generation quarks, $a_{tb}$,
which are important for the decay of the $\PWpr$ boson, are equal.
For given values of $a^L$ and $a^R$, the distributions are obtained by
combining the four signal samples according to Eq.~(\ref{eq:xsec}).

\begin{figure}[htbp]
\begin{center}
\includegraphics[width=\cmsFigWidth]{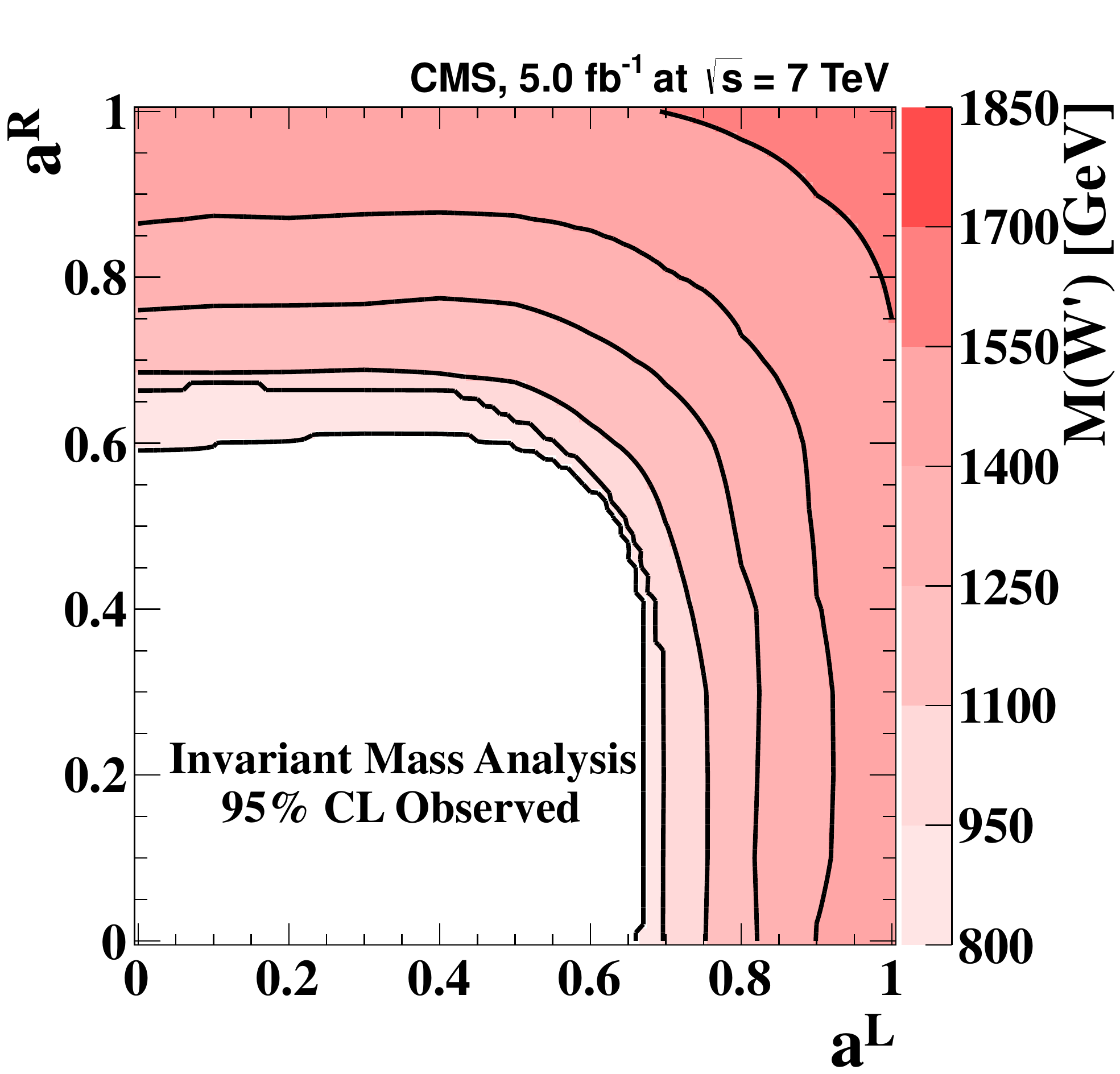}
\includegraphics[width=\cmsFigWidth]{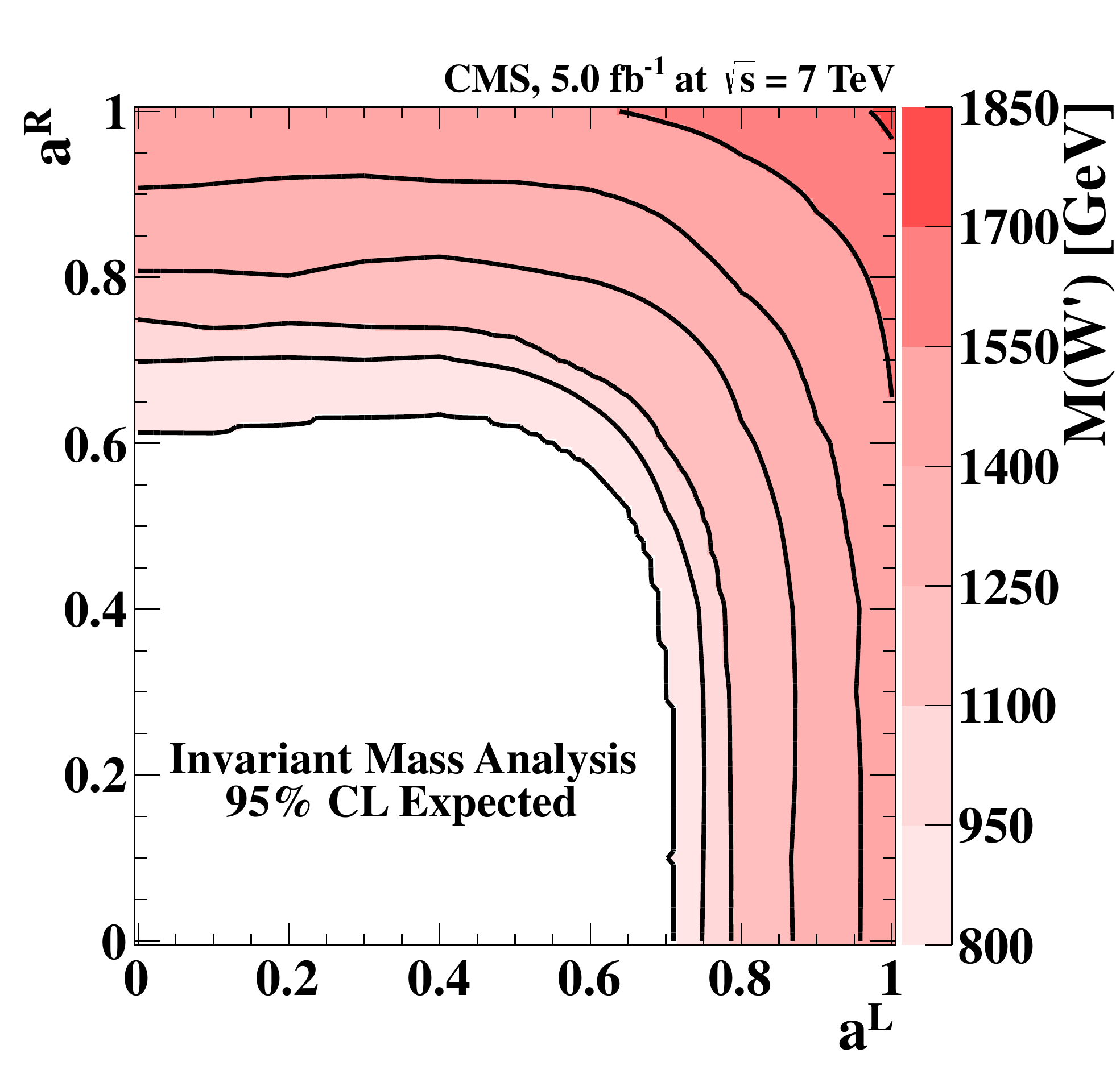}
\caption{
 Contour plots of  $M(\PWpr)$ in the ($a^L,a^R$) plane
at which the 95\% CL upper cross section limit equals the predicted
cross section for the combined $\Pe,\Pgm$+jets sample.
The \cmsLeft\ (\cmsRight) panel is for observed (expected) limits.
The color-scale axis shows the ${\PWpr}$ mass in \GeV.
The dark lines represent equispaced contours of ${\PWpr}$ mass
at 150\GeV intervals.
\label{fig:disc_limit}}
\end{center}
\end{figure}

We vary both $a^L$ and $a^R$ between 0 and 1 in steps of 0.1, for a series of values of the mass of the
{\PWpr} boson. Templates of the reconstructed {\PWpr} invariant mass distributions are generated
for each set of  $a^L$, $a^R$,  and $M({\PWpr})$ values
by weighting the events from the four simulated samples,
as described in Sec.~\ref{sec:MCmodel}, according to Eq.~\ref{eq:xsec}.
For each of these combinations of $a^L$, $a^R$,  and $M({\PWpr})$, we determine the
expected and observed 95\% CL upper limits on the cross section.
We then assume values for $a^L$, and $a^R$,
and interpolate the cross section limit in  the mass value.
Figure~\ref{fig:disc_limit} shows the contours
for the {\PWpr} boson mass  in the
($a^L,a^R$) plane for which the cross section limit equals the predicted cross section.
For each contour of $\PWpr$ mass, combinations of the couplings $a^R$ and $a^L$ above and to the right of the curve are excluded
The contours are obtained using the {\PWpr} invariant mass distribution.
For this analysis, we make the conservative assumption that $M_{\nu_R} \ll M_{{\PWpr}}$.
The observed lower limit on the mass of the {\PWpr} boson with coupling to purely left-handed fermions and
with couplings to both left- and  right-handed fermions with equal strength is listed in Table~\ref{tab:masslim}.

\begin{table*}[!h!tb]
\begin{center}
\small
{\topcaption{Observed lower limit on the mass of the ${\PWpr}$ boson.
For $\PWpr$ with right-handed couplings, we consider two cases for
the right-handed neutrino: $ M_{\nu_R} > M_{{\PWpr}}  $
and $ M_{\nu_R} \ll M_{{\PWpr}} $.
}
\label{tab:masslim}
        }
\footnotesize
\begin{tabular}{l|cc|c|c} \hline \hline
 & \multicolumn{2}{c|}{$(a^L,a^R)=(0,1)$} & $(a^L,a^R)=(1,0)$ & $(a^L,a^R)=(1,1)$ \\
Analysis & $M_{\nu_R} > M_{{\PWpr}}$ & $M_{\nu_R} \ll M_{{\PWpr}}$ & $M_{\nu_R} \ll M_{{\PWpr}}$ & $M_{\nu_R} \ll M_{{\PWpr}}$  \\ \hline
BDT & 1.91 \TeV &1.85 \TeV & --- & --- \\
Invariant Mass & --- & --- & 1.51 \TeV & 1.64 \TeV \\
\hline \hline
\end{tabular}
\normalsize
\end{center}
\end{table*}

\section{Summary}

A search for \PWpr~boson production in the $\cPqt\cPqb$ decay
channel has been performed in pp collisions at $\sqrt{s}=7$\TeV  using  data corresponding to an integrated luminosity
of \usedLumiWprime collected during 2011 by the CMS experiment at the LHC.
Two analyses have searched for $\PWpr$ bosons, one uses the
reconstructed $\cPqt\cPqb$ invariant mass analysis to search
for {\PWpr} bosons with arbitrary combinations of left-~and right-handed couplings
while a multivariate analysis is optimized for the search of {\PWpr} bosons with
purely right-handed couplings.
No evidence for \PWpr~boson production is found and 95\% CL
upper limits on the production cross section times branching ratio are set
for arbitrary mixtures of couplings to left-~and right-handed fermions.
Our measurement is compared to the theoretical prediction for the nominal value of the
cross section to determine the lower limits on the mass of the \PWpr.
For \PWpr~bosons with right-handed couplings to fermions
a limit of 1.85  (1.91)\TeV is established when $M_{{\nu_R}} \ll M_{\PWpr}$ ($M_{\nu_R} > M_{{\PWpr}}$).
This limit also applies for \PWpr~bosons with left-handed couplings to fermions when no interference with SM {\PW} boson is included.
In the case of interference, and for $M_{{\nu_R}} \ll M_{\PWpr}$, the limit obtained is $M_{\PWpr}> 1.51$\TeV for
purely left-handed couplings and
 $M_{\PWpr}>1.64 $ \TeV  if both left-~and right-handed couplings are present.

For the first time using the LHC data, constraints on the \PWpr~gauge couplings for a set of
left-~and right-handed coupling combinations have been placed.
These results represent a significant improvement over previously published limits
in the case of the $\cPqt\cPqb$ final state.

\section*{Acknowledgements}
We congratulate our colleagues in the CERN accelerator departments for the excellent performance of the LHC machine. We thank the technical and administrative staff at CERN and other CMS institutes, and acknowledge support from: BMWF and FWF (Austria); FNRS and FWO (Belgium); CNPq, CAPES, FAPERJ, and FAPESP (Brazil); MES (Bulgaria); CERN; CAS, MoST, and NSFC (China); COLCIENCIAS (Colombia); MSES (Croatia); RPF (Cyprus); MEYS (Czech Republic); MoER, SF0690030s09 and ERDF (Estonia); Academy of Finland, MEC, and HIP (Finland); CEA and CNRS/IN2P3 (France); BMBF, DFG, and HGF (Germany); GSRT (Greece); OTKA and NKTH (Hungary); DAE and DST (India); IPM (Iran); SFI (Ireland); INFN (Italy); NRF and WCU (Korea); LAS (Lithuania); CINVESTAV, CONACYT, SEP, and UASLP-FAI (Mexico); MSI (New Zealand); PAEC (Pakistan); MSHE and NSC (Poland); FCT (Portugal); JINR (Armenia, Belarus, Georgia, Ukraine, Uzbekistan); MON, RosAtom, RAS and RFBR (Russia); MSTD (Serbia); SEIDI and CPAN (Spain); Swiss Funding Agencies (Switzerland); NSC (Taipei); TUBITAK and TAEK (Turkey); NASU (Ukraine); STFC (United Kingdom); DOE and NSF (USA).
\bibliography{auto_generated}   % will be created by the tdr script.

\cleardoublepage \appendix\section{The CMS Collaboration \label{app:collab}}\begin{sloppypar}\hyphenpenalty=5000\widowpenalty=500\clubpenalty=5000\textbf{Yerevan Physics Institute,  Yerevan,  Armenia}\\*[0pt]
S.~Chatrchyan, V.~Khachatryan, A.M.~Sirunyan, A.~Tumasyan
\vskip\cmsinstskip
\textbf{Institut f\"{u}r Hochenergiephysik der OeAW,  Wien,  Austria}\\*[0pt]
W.~Adam, T.~Bergauer, M.~Dragicevic, J.~Er\"{o}, C.~Fabjan\cmsAuthorMark{1}, M.~Friedl, R.~Fr\"{u}hwirth\cmsAuthorMark{1}, V.M.~Ghete, J.~Hammer, N.~H\"{o}rmann, J.~Hrubec, M.~Jeitler\cmsAuthorMark{1}, W.~Kiesenhofer, V.~Kn\"{u}nz, M.~Krammer\cmsAuthorMark{1}, D.~Liko, I.~Mikulec, M.~Pernicka$^{\textrm{\dag}}$, B.~Rahbaran, C.~Rohringer, H.~Rohringer, R.~Sch\"{o}fbeck, J.~Strauss, A.~Taurok, P.~Wagner, W.~Waltenberger, G.~Walzel, E.~Widl, C.-E.~Wulz\cmsAuthorMark{1}
% ...
\vskip\cmsinstskip
\dag:~Deceased\\
1:~~Also at Vienna University of Technology, Vienna, Austria\\
% ...

\end{sloppypar}
\end{document}